\documentclass[
aps,%
10pt,%
final,%
notitlepage,%
oneside,%
onecolumn,%
nobibnotes,%
nofootinbib,
superscriptaddress,%
noshowpacs,%
centertags]%
{revtex4}
\usepackage{graphicx,amssymb,color}
\usepackage{amsmath}

\begin{document}

%
\title{On the distribution function of {suprathermal} particles at collisionless shocks}
\author{\firstname{Bojan} \surname{Arbutina}}
\email[]{arbo@matf.bg.ac.rs}
%
\affiliation{Department of Astronomy, Faculty of Mathematics,
University of Belgrade, Studentski trg 16, 11000 Belgrade, Serbia}
\author{\firstname{Vladimir} \surname{Zekovi\'c}}
\email[]{vlada@matf.bg.ac.rs}
\affiliation{Department of Astronomy, Faculty of Mathematics,
University of Belgrade, Studentski trg 16, 11000 Belgrade, Serbia}
\affiliation{Department of Astrophysical Sciences, Princeton University, Princeton, NJ 08544, USA}

%
\begin{abstract}
{\bf Abstract.} The departure of particle distributions from the Maxwellian is commonly observed in space plasmas. These non-Maxwellian distributions which are typical for plasmas that are not in thermal equilibrium, can be modeled with $\kappa$-distribution. Kinetic simulations of quasi-parallel collisionless shocks show that proton distribution is a composite of thermal, suprathermal, and non-thermal parts. By using particle-in-cell shock simulations, we show that $\kappa$-distribution adequately fits thermal and suprathermal parts together, as {a single} continuous distribution in early proton spectra. {We derive suprathermal proton distribution directly from the generalized entropy of non-extensive statistical mechanics, and show that thermal and suprathermal populations are both naturally embedded in $\kappa$-distribution.} We find that the index $\kappa$ of the distribution increases with the distance from the shock, following the decrease in suprathermal part. {The non-equilibrium plasma distribution which is continuously being enriched with suprathermal particles at the reforming shock barrier, reaches the thermal equilibrium in the far downstream. The suprathermal part completely fades there, and the shape of proton distribution becomes a Maxwellian from which directly emerges a power-law.}
\end{abstract}

\maketitle
%

%
%
%
%
%
%


\section{Introduction}

{The diffusive} shock acceleration (DSA) theory as a promising mechanism of acceleration of particles at interstellar collisionless shocks, up to the cosmic-ray (CR) energies, was proposed independently by Axford et al. (1977), Krymsky (1977) and Blandford \& Ostriker (1978) and Bell (1978). Besides CRs observed at Earth that show a characteristic power-law spectral form predicted by the theory (but modified for effects of transport in the Galaxy), indirect evidence for DSA power-law spectra (for electrons, at least) come from e.g. radio observations of supernova remnants and other astrophysical synchrotron emitting sources (see e.g. Arbutina 2017, {Vink 2020}). Nevertheless, little is {still known} about the particle spectra at lower energies (momenta). An important insight about the distribution function of particles at collisionless shocks in thermal, suprathermal and non-thermal regime can be provided by particle-in-cell (PIC) simulations. PIC and hybrid simulations (in which electrons are treated as a fluid) do, indeed, show the power-law spectra a'la Bell (1978) at {higher} momenta (Caprioli \& Spitkovsky 2014a,b,c, Caprioli et al. 2015). In this {paper}, {for the sake of completness, we  rederive} the DSA theory to account for particles of lower momenta, introduce the $\kappa$-distribution as a plausible function for thermal and suprathermal parts of the spectra, and compare the theoretical spectrum with simulations.

\section{ACCELERATED PARTICLES}

Of all of the mentioned approaches to DSA, Bell's (1978) microscopic approach is probably the most intuitive, since it tries to explain what is happening to individual particles in the process of acceleration. The idea is that sufficiently energetic particles can cross and recross the shock from downstream to upstream and vice verse multiple times, scattering of turbulence and magnetic instabilities present, {and, in each cycle,} can gain energy i.e. momentum $\frac{\Delta p}{p} \approx \frac{4}{3}\frac{u_1-u_2}{v}$ (as in 1$^\mathrm{st}$ order Fermi acceleration). Bell (1978) argued that the probability of a particle engaged in DSA cycles to be advected downstream is $\mathcal{P} = \frac{4u_2}{v}$, and consequently, the probability to cross back to upstream and stay in DSA is
\begin{equation}
\label{eq1} \mathcal{P}_B = 1 - \frac{4u_2}{v},
\end{equation}
where $u_2$ is the velocity of the downstream plasma as seen from the shock frame, and $v$ is a particle velocity in the plasma frame. We will rederive this probability inspired by approach found in {Vietri (2008) and} Blasi (2012). Assuming a monoenergetic distribution of particles with number density $dN = 4 \pi p^4 f(p) dp$,  the flux density in the shock frame $\varphi = \frac{1}{2} \int v' dN' d\mu'$ can be written as
\begin{equation}
\label{eq2} \varphi = \frac{1}{2} \int _{-1} ^1 \frac{u_2 +\mu v}{1+\frac{u_2 v \mu}{c^2}} dN d\mu,
\end{equation}
where $v$ and $\mu = \cos \theta$ are now measured in the downstream frame, and $f(p)$ is assumed do be isotropic. As seen from this frame, in which the downstream plasma is at rest, the shock is moving with velocity $-u_2$, so the particles with $v_x = v \mu >-u_2$ i.e. $\mu > -\frac{u_2}{v}$ are able to cross {the shock and return upstream}. We can then define, similarly to Blasi (2012), flux densities
\begin{equation}
\label{eq3}  \mathcal{P}_B = \frac{|\varphi _{out}|}{\varphi _{in}},\ \ \varphi _{in} = \int _{-\frac{u_2}{v}} ^1 d\varphi ,\ \ \varphi _{out} =\int _{-1} ^{-\frac{u_2}{v}} d\varphi,
\end{equation}
and obtain a quite general expression for Bell's probability
\begin{equation}
\label{eq4} \mathcal{P}_B = \frac{\Big|1-\frac{u_2}{v} - \frac{c^2}{u_2 v}\Big(1- \frac{u_2^2}{c^2} \Big)\ln\Big| \frac{1-\frac{u_2^2}{c^2}}{1-\frac{u_2 v}{c^2}}\Big|\Big|}{1+\frac{u_2}{v} - \frac{c^2}{u_2 v}\Big(1- \frac{u_2^2}{c^2} \Big)\ln\Big| \frac{1+\frac{u_2 v}{c^2}}{1-\frac{u_2 ^2}{c^2}}\Big|}.
\end{equation}
For non-relativistic shocks, i.e. $u_2 \ll c$, one can check that the last expression reduces to
\begin{equation}
\label{eq5} \mathcal{P}_B = \Bigg( \frac{1-\frac{u_2}{v}}{1+\frac{u_2}{v}}\Bigg) ^2 \approx 1- \frac{4u_2}{v}.
\end{equation}

Armed with general Bell's probability we are {able to derive} an approximate distribution function for non-thermal particles even at lower momenta. We are assuming all the time that we are in a test-particle regime, i.e not considering CRs backreaction and modification of the shock (non-linear DSA). For two consequent DSA cycles $k$ and $k+1$, we will have $N_{k+1}/N_k = \mathcal{P}_B (k)$, where $N_i=N(p>p_i)$ is the cumulative number of particles per unit volume with momentum larger than $p_i$, and $p_{k+1}/p_k \approx 1 +\frac{4}{3} \frac{u_1-u_2}{v_k}$. We will in the following omit index $k$ in $N_k \rightarrow N$, $v_k \rightarrow v$, and, similarly to Caprioli et al.~(2015), define cumulative number change and momentum gain $\mathcal{G}$ through:
\begin{equation}
\frac{\Delta N}{N} = \rightarrow \frac{dN}{N} = -\mathcal{P} = \mathcal{P}_B -1 = \frac{-\frac{4u_2}{v}}{(1+u_2/v)^2},
\end{equation}
\begin{equation}
\frac{\Delta p}{p} = \rightarrow \frac{dp}{p} = \mathcal{G} = \frac{4(R-1)}{3}\frac{u_2}{v},
\end{equation}
where $R=u_1/u_2$ is the compression ratio ($u_1$ is shock velocity, as observed from laboratory frame). From the above equations, we have
\begin{equation}
\frac{d\ln N}{d\ln p} = -\frac{\mathcal{P}}{\mathcal{G}},
\end{equation}
from which, by using $dN = - 4 \pi p^2 f dp$, one can derive
\begin{equation}
\frac{d\ln \Big(\frac{\mathcal{G}}{\mathcal{P}}f\Big)}{d\ln p} = -\Big(\frac{\mathcal{P}}{\mathcal{G}}+3\Big) = -3 \Big( 1+ \frac{1}{R-1}\frac{1}{(1+ \frac{u_2}{v})^2} \Big).
\end{equation}
This equation can be integrated {to give the solution (for non-relativistic shocks):}
\begin{equation}
\label{sol}
f(p) = \frac{3N_{CR}}{4\pi (R-1) p_{inj}^3}  \Big(1+\frac{u_2}{v_{inj}}\Big)^{\frac{3}{R-1}} \Bigg(\frac{p}{p_{inj}}\Bigg)^{-\frac{3R}{R-1}} \Big(1 +  \frac{u_2}{v}\Big)^{-\frac{2R+1}{R-1}}   e^{\frac{3  u_2}{R-1}(\frac{1}{v_{inj}+u_2} - \frac{1}{v +u_2})},
\end{equation}
{where $N_{CR}$ is the total number of CRs, and $p_{inj}$ is some injection momentum. For $p \gg m u_2$ Eq. (\ref{sol}) gives the well-known dependence $f(p) \propto (p/p_{inj})^{-3R/(R-1)}$ i.e. $f(p) \propto p^{-4}$ for the standard compression $R=4$. Deviation from the power-law i.e. the modification at lower momenta may not be relevant because specularly-reflected particles never become isotropic in the downstream, violating Bell's initial assumption.
Nevertheless, in Fig. 2 we have also plotted this distribution assuming that a fraction of particles at lower momenta complying with Bell's assumption, conditionally speaking, belongs to the non-thermal, rather than suprathermal population.}

\section{PRE-ACCELERATED AND THERMAL PARTICLES}

Because of the self-reforming behaviour, the collisionless shocks are expected to produce {non-equilibrium plasmas}. Indeed, the non-Maxwellian particle distributions are observed at the shock of supernova remnants (Raymond et al. 2010). This is also {found} in kinetic (PIC or hybrid) simulations. Caprioli et al. (2015) suggested that the plasma distribution downstream of the shock can be described by the sum of thermal, suprathermal, and non-thermal components, that we will denote with $f_T$, $f_S$ and $f_N$, respectively. As described in the minimal model for ion injection (Caprioli et al. 2015), while most of the (thermal) ions will be advected and isotropized after crossing the shock, some ions can gain extra energy by performing a few gyrations while drifting along the shock surface (the shock drift acceleration - SDA). These suprathermal ions, if continuing SDA or micro-DSA (Zekovi\'c \& Arbutina 2019) cycles, can later provide the seed particles for the standard DSA mechanism.

By including the finite duty cycle of a reforming shock barrier, one can easily model the suprathermal transition observed in the particle spectra (Caprioli et al. 2015). This can be accomplished by assuming a modified probability for a particle to cross to upstream i.e. stay in the cycles, e.g. $\mathcal{P}_A \cdot \mathcal{P}_B$, where $\mathcal{P}_A$ can be thought as a probability for a particle to pass through the shock of some finite tickness, not being halted or reflected back downstream. In the SDA case, Caprioli et al. (2015) assumed the constant total probability for a particle to overcome the periodically reforming shock barrier with the duty cycle, i.e. $\mathcal{P} = 1- \mathcal{P}_A \mathcal{P}_B = 0.75$. This implies that roughly 75 percent of particles would be thermalized, while the remaining 25 percent would become suprathermal or non-thermal. If again
\begin{equation}
\frac{d\ln S}{d\ln p} = -\frac{\mathcal{P}}{\mathcal{G}},
\end{equation}
where now $S = S(>p)$ is the cumulative number of suprathermal particles, by using $dS = - 4 \pi p^2 f_S dp$, one can derive
\begin{equation}
f_S = \frac{3N_{ST} \mathcal{P}}{16\pi (R-1) m u_2 p^2}  {e^{\frac{3 \mathcal{P}}{4(R-1)m u_2}(p_{min} - p)}} ,
\end{equation}
where $N_{ST}$ is the total number of suprathermal particles, and $p_{min}$ is the momentum at which particles enter SDA. {We assumed that the particles are non-relativistic, i.e. $p = mv$.}
One can see that the number of suprathermal particles per unit momentum is actually an exponential function. For simplicity, $f$ can be modeled as the sum, $f_T+f_S+f_N$, where $f_N$ is taken to be in the standard power-law form
\begin{equation}
f_N =  \frac{3N_{CR}}{4\pi p_{inj}^3 (R-1)} (p/p_{inj})^{-3R/(R-1)}.
\end{equation}
and the thermal distribution is a Maxwellian\footnote{If particles are relativistic, one has the Maxwell-J\"utter distribution (see Synge 1957)
\begin{equation}
f_T = \frac{N_0 }{4 \pi m^3 c^2 \Theta K_2(1/\Theta)} e^{-\frac{\sqrt{1+\frac{p^2}{m^2c^2}}}{\Theta}},
\end{equation}
where $\Theta = kT/(mc^2)$, $T$ is thermodynamic temperature and $K_2(x)$ the modified Bessel function of the 2$^\mathrm{nd}$ order (see Abramowitz \& Stegun 1972).}
\begin{equation}
f_T =  \frac{N_0}{(2\pi m kT)^{3/2}} e^{-\frac{p^2}{2mkT}}.
\end{equation}
{The distributions} $f_S$ and $f_N$ of course start at some initial momenta $p_{min}$ and $p_{inj}$, respectively.

The minimal model tries to describe suprathermal and non-thermal particle distributions, however it cannot describe the {thermal} maximum in the spectra {by a suprathermal distribution itself.} One could also try to describe thermal and suprathermal particle distributions with one {non-equilibrium} distribution $f = f_{NE}$. In order to model both these components as different features of a single, non-stationary plasma, {in} Arbutina \& Zekovi\'c (2020) {we introduced} a $\kappa$-distribution. It is shown by Livadiotis (2017) that the state of a plasma which has not reached the thermodynamic equilibrium, can be characterized by this distribution, with index $\kappa$ being a free parameter serving as sort of a measure of non-equilibrium. Such distributions are common to the space plasmas (Livadiotis \& McComas 2011), {including the Solar wind (Martinovi\'c 2016)}.
{Non-relativistic $\kappa$-distribution may be written as
\begin{equation}
f_{NE} = \frac{N_0}{(\pi \kappa p_0^2)^{3/2}} \frac{\Gamma(\kappa +1)}{\Gamma(\kappa -\frac{1}{2})} \frac{1}{\Big[ 1+ \frac{p^2}{\kappa p_0^2} \Big]^{\kappa +1}},\ \ \ p_0^2 = 2 mkT
\end{equation}
which for $\kappa  \to \infty$ tends to become a Maxwellian. Note that $T$ is not the usual thermodynamic temperature if plasma is out of equilibrium.

In order to understand theoretically $\kappa$-distribution one can introduce the Tsallis entropy, which is shown (Livadiotis 2017, Tsallis 2017) to be the generalization of a Boltzmann-Gibbs (BG) entropy from which Maxwellian distribution is derived. This generalization naturally arises if instead of a standard relation of BG entropy $S_{\rm BG} = -k_{\rm B} \sum_{i=1}^{\rm W} p_i \ln p_i$ ($p_i$ is the probability of state $i$) one assumes the most general relation $S = k_{\rm B} \sum_{i=1}^{\rm W} f(p_i)$, where $f(p_i)$ is some arbitrary function of $p_i$. By following the principle of maximization:
\begin{equation}
    \frac{\partial}{\partial p_j} S(p_1, p_2,...,p_{\rm W}) + \lambda_1 - \lambda_2 \epsilon_j = 0,
\end{equation}
and applying the energy and entropy additivity among two parts of the system:
\begin{equation}
    \epsilon_{ij}^{\rm A+B} = \epsilon_{i}^{\rm A} + \epsilon_{j}^{\rm B},\ S_{ij}^{\rm A+B} = S_{i}^{\rm A} + S_{j}^{\rm B},
\end{equation}
the general (Tsallis) entropy is derived (Livadiotis 2017):
\begin{equation}
    S =  k_{\rm B} \frac{1}{q-1} \sum_{i=1}^{\rm W} (p_i - p_i^q),\ \ \ q = 1 + \frac{1}{\kappa}.
\end{equation}

This entropy can be expanded to a series of infinite $(q-1)$ terms as:
\begin{equation}
    S =  -k_{\rm B} \sum_{i=1}^{\rm W} p_i \ln p_i - k_{\rm B} \sum_{n=2}^\infty \left[ \frac{1}{n!} (q-1)^{n-1} \cdot \sum_{i=1}^{\rm W} p_i \ln^n p_i \right].
\end{equation}

From PIC runs (see the next section) we find that $q \sim 1.2$, so we keep only the BG term and the $n=2$ term in the series, as higher order terms becomes negligible:
\begin{equation}
    S \cong -k_{\rm B} \sum_{i=1}^{\rm W} p_i \ln p_i - k_{\rm B}  \frac{1}{2} (q-1) \cdot \sum_{i=1}^{\rm W} p_i \ln^2 p_i.
\end{equation}
Since BG term describes the thermal component, the $n=2$ term then must correspond to the departure from the equilibrium, which is caused by the presence of suprathermal particles. By using Eq.(17) with the $n=2$ term, we derive this non-equilibrium departure from the thermal distribution function:
\begin{equation}
    f_{\Delta} \sim a \cdot e^{b \frac{\sqrt{\kappa}}{m u_2} (p_{min} - p)},
\end{equation}
where $a$ and $b$ are constants.

We find that $f_{\Delta}$ has very similar dependence on $p$ as the suprathermal distribution $f_S$ given by Eq.(12) which is derived by assuming the constant escape probability (as in the minimal model of Caprioli et al. 2015). The shape of the distribution $f_{\Delta}$ also finely agrees with the suprathermal distribution function given in Caprioli et al. (2015). The difference between $f_{\Delta}$ and $f_S$ due to the extra dependence $1/p^2$ in $f_S$, can be directly overcome by introducing a weak logarithmic dependence:
\begin{equation}
    \sim 0.75 - \frac{c}{( p - p_{min} )} \ln \left(\frac{p}{p_{\rm sh}} \right)^{2}
\end{equation}
(where $c$ is a constant) into the escape probability $\mathcal{P}$ in Eq.(12) which is justified as $p \sim 1 - 2\ p_{\rm sh}$ in the range of suprathermal momenta. Also, by only adjusting the parameter $b$ in Eq.(22), the distributions $f_{\Delta}$ and $f_S$ can completely overlap.

Therefore, the theory implies that thermal and suprathermal distributions are both naturally embedded into single, non-equilibrium plasma distribution function. The non-stationary reforming shock barrier thus acts as a generator of the non-equilibrium states in plasma, that result in $\kappa$-distribution. While this unique distribution is commonly interpreted as a superposition of thermal and suprathermal components, it is actually composed of particles whose energies are set by a single physical process.

}

\section{PIC simulations}

We run the long 1D particle-in-cell simulation of an initially parallel collisionless shock by using the PIC code TRISTAN-MP (Spitkovsky 2005). The parameters of the run are shown in Table 1. As it is commonly done, we initiate the shock by reflecting the plasma beam from the left wall of a simulation domain. We use the expanding simulation box which enlarges ahead of the shock, as the moving plasma injector reaches the right wall of the domain. By this, we are able to significantly extend the evolution of a shock with the given mass ratio. We resolve the electron skin depth ($c/\omega_{pe}$) with 10 cells, and each cell initially contains 8 particles (4 electrons and 4 ions). The noise is reduced by filtering particle contribution to the current 32 times per timestep. As it is shown in Sironi \& Spitkovsky (2011), the mass ratio $m_i/m_e = 16$ is large enough to separate the ion and electron scales. By the simulation end time $t \sim 2220~ \omega_{ci}^{-1}$, the shock enters the quasi-equilibrium stage. The phase space, and density and field profiles are given in Fig. 1.

\begin{table}[h!]
    \caption{The parameters of the run from left to the right: ion-to-electron mass ratio, ion magnetization (the ratio of magnetic to kinetic energy density), the shock velocity in the lab frame, the Alfven-Mach and sonic Mach numbers, and the simulation end time, respectively.
}
     \label{table1}
     \vskip.25cm
\centerline{\begin{tabular}{cccccc} \hline
$m_i/m_e$ & $\sigma_i$ & $u_1/c$ & $M_A$ & $M_S$ & $t\ [\omega_{ci}^{-1}]$ \\
\hline
16 & $6\times 10^{-4}$ & 0.4 & 16 & 35 & 2220 \\
\hline
\end{tabular}
}
\end{table}

\begin{figure}[ht!]
  \centering
  \includegraphics[bb = 0 0 730.8 544.32, width=\textwidth,keepaspectratio]{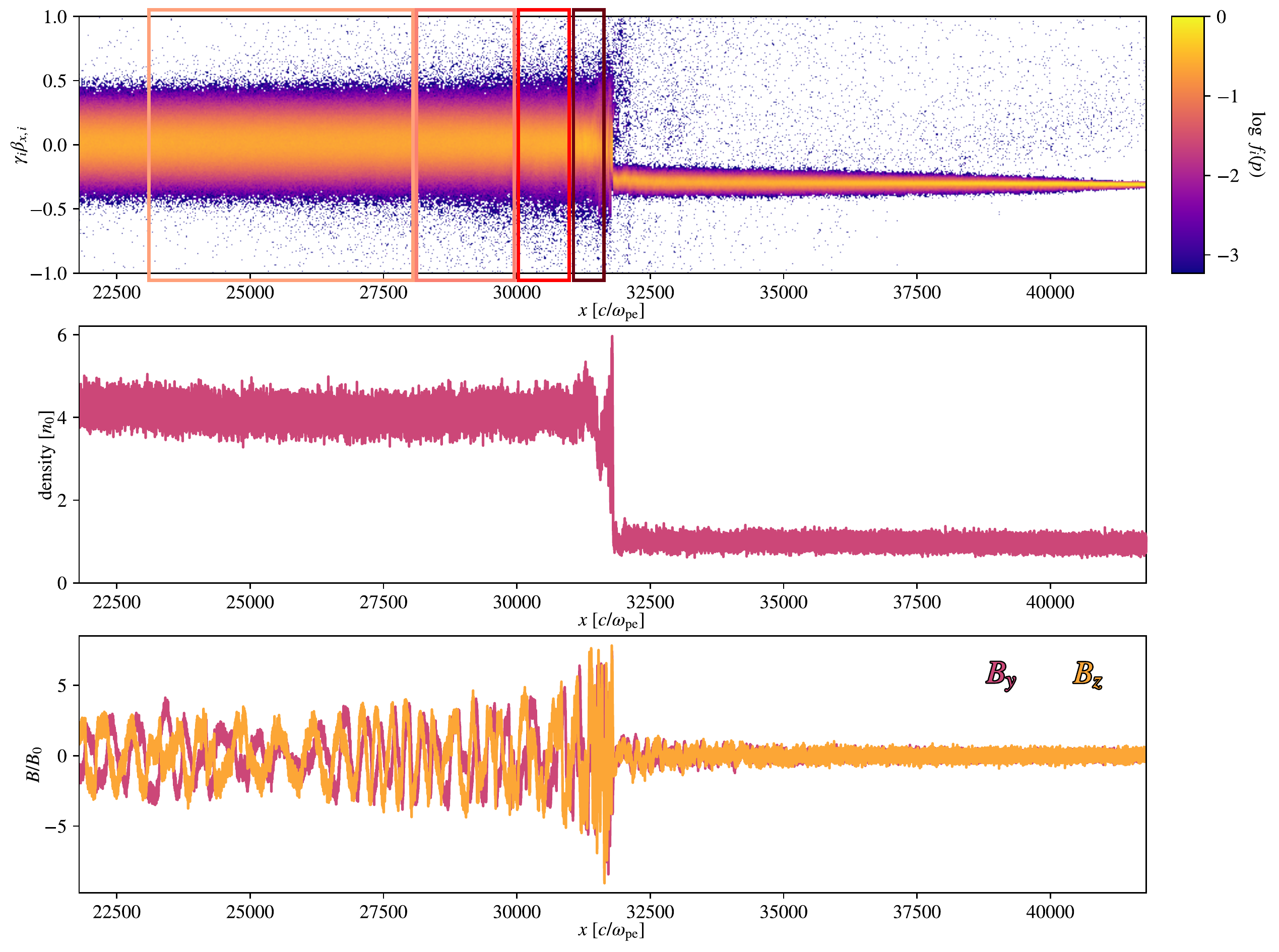}
\caption{The longitudinal phase space, density profile, and transverse $B_y$ and $B_z$ magnetic field profiles, given at the simulation end time. {The regions from which ion distributions are measured and plotted in Figs. 2 and 3, are shown in the top graph.}}
\end{figure}

At the beginning of the run, the Weibel-type instability (Weibel 1959) grows fastest than Alfv\`enic type instabilities (Crumley et al. 2019). Once the wave driven by the resonant instability (Zekovi\'c 2019) grows to $\sim B_0$, the shock (re)formation is further mediated by Alfv\`enic modes (as shown in Fig. 1). The current of reflected ions seeds the upstream wave via non-resonant streaming instability (Bell's, or CR streaming instability; Bell 2004, Amato \& Blasi 2009) which pre-accelerate particles at the shock interface through the SDA mechanism (Caprioly et al. 2015). The particles that escape upstream, become eligible to enter the DSA process.

\begin{figure*}[t!]
  \centering
  \includegraphics[bb = 0 0 1079 426, width=0.9\textwidth,keepaspectratio]{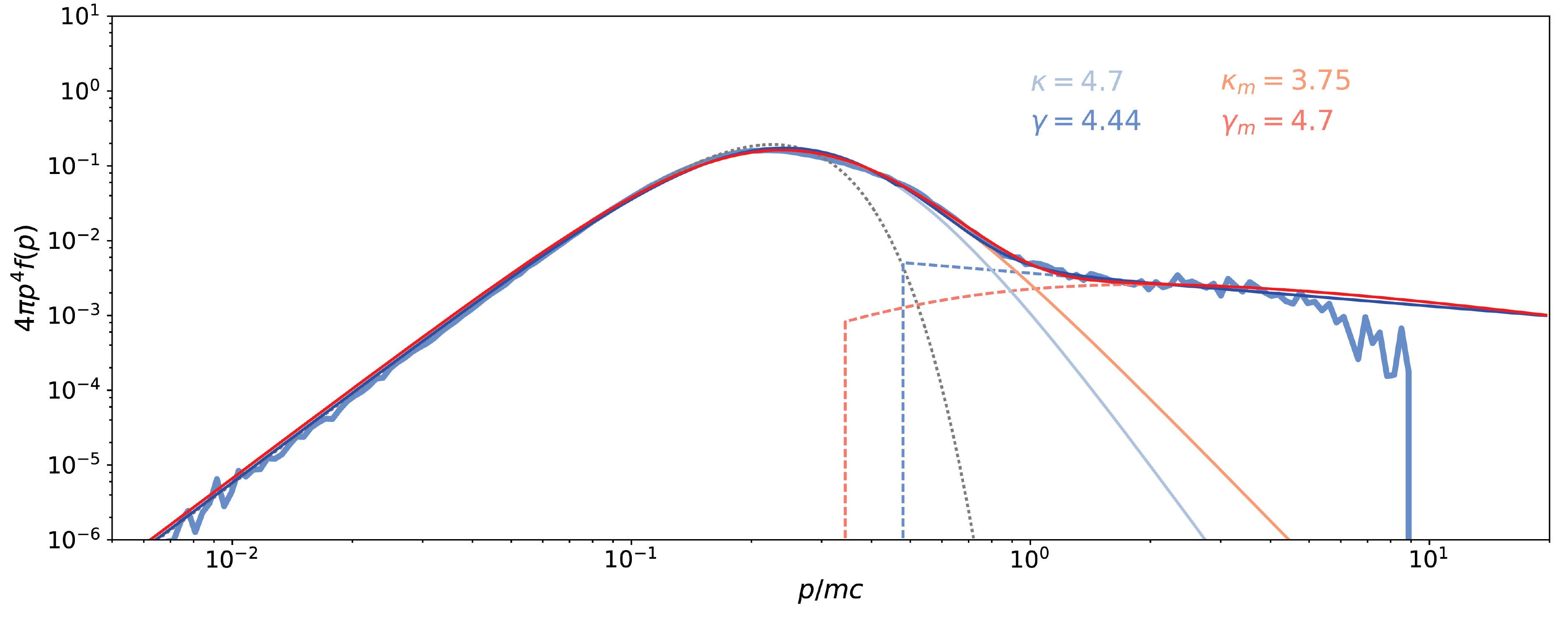}
\caption{The post-shock ion spectrum {(the sampling region is marked by the brown square in Fig. 1).} The dotted grey line denotes Maxwellian fit; the lighter blue and red lines correspond to $\kappa$-distribution fits for the cases of pure and modified power-laws, respectively; the {dashed} lines correspond to the two power-laws; the strong lines are the best $\kappa$ + power law fits; the data is plotted as a thick, light-blue curve. The measured parameter $\kappa$ and spectral slope  $\gamma$ of the momentum distribution $f(p) \propto p^{-\gamma}$ are given for the two cases in the upper right corner.
}
\end{figure*}

The initial particle spectrum in PIC simulations is clearly not Maxwellian. This can also be seen in the spectral plot in Caprioli \& Spitkovsky (2014a). We here report that the thermal and suprathermal components in ion spectra may correspond to {the} plasma that is in {a} transient state, which is ideally fitted by the $\kappa$-distribution function (see Fig. 2). We fit the whole momentum spectrum by the sum of $\kappa$ and power-law distributions{, where we at first find the best fit for $\kappa$-distribution. In the next step, we search for the best value of the momentum $p_{inj}$ at which ions are injected into DSA, so that the $\kappa$ + power-law curve fits the whole spectrum (as shown in Fig. 2).} We present the two {fitting combinations}, one where we use the pure power-law given by {Eq.(13)}, and the other which is modified at lower momenta according to {Eq.(10). In both cases, we use Eq. (16) to fit the thermal + suprathermal parts as one unique non-equilibrium plasma distribution.} From the best fit we get that the momentum at which ions enter DSA is $p_{inj} \sim 1.5-2~p_{sh} = 6-8~m u_2$. {It is lower than the injection momentum given in the minimal model of Caprioli et al. (2015) which is defined as a momentum that particles need to reach by performing few SDA cycles in order to enter DSA. Here, we define $p_{inj}$ as a momentum at which only a small fraction of ions that will eventually reach high energies (whose amount is less than $4 \%$ according of Caprioli et al. 2015) enter their first cycle of acceleration -- the momentum gained after the first specular reflection. Since these ions originate from the tail of $\kappa$-distribution (to the right of $p_{inj}$) {we do not expect exactly $\delta$-function injection into DSA, but rather a steep power-law injection (see Arbutina 2017), meaning that $p_{inj}$ approximately} represents the lowest injection momentum. By the power-laws shown in Fig. 2, we therefore model only the population of ions that escaped the reforming shock barrier and populated the non-thermal tail in the process of DSA. For those ions, an energy gain per each cycle is that of DSA (the same as in Caprioli et al. 2015) and their escape probability is the Bell's probability (given by Eq. (5)) from the first reflection (even during SDA). This confirms our previous assumptions that SDA governed by the instabilities at quasi-parallel shocks, can also be described by a process which is physically the same as DSA (Zekovi\'c \& Arbutina 2019). As scattering centers during the first few reflections (SDA cycles) are provided by the micro-structure of a reforming shock barrier itself, we have earlier named this process a $\mu$-DSA (Zekovi\'c \& Arbutina 2019).

In the previous section, we argued that the distribution of suprathermal particles (induced by a constant probability of escape in SDA mechanism) can be derived directly from the second order term in Tsallis entropy, while Maxwellian distribution is derived from the first order term, and higher order terms produce only smaller corrections. Therefore, we believe that the reforming barrier induces a non-equilibrium states in the shocked plasma. Instead of having independent distributions for each population, the out-of-equilibrium plasma can easily be modeled by a single particle distribution that encompasses both, thermal and suprathermal population of ions produced in a non-stationary shock reformation process.}

\begin{figure}[t!]
  \centering
  \includegraphics[bb = 0 0 1080 520.56, width=0.9\textwidth,keepaspectratio]{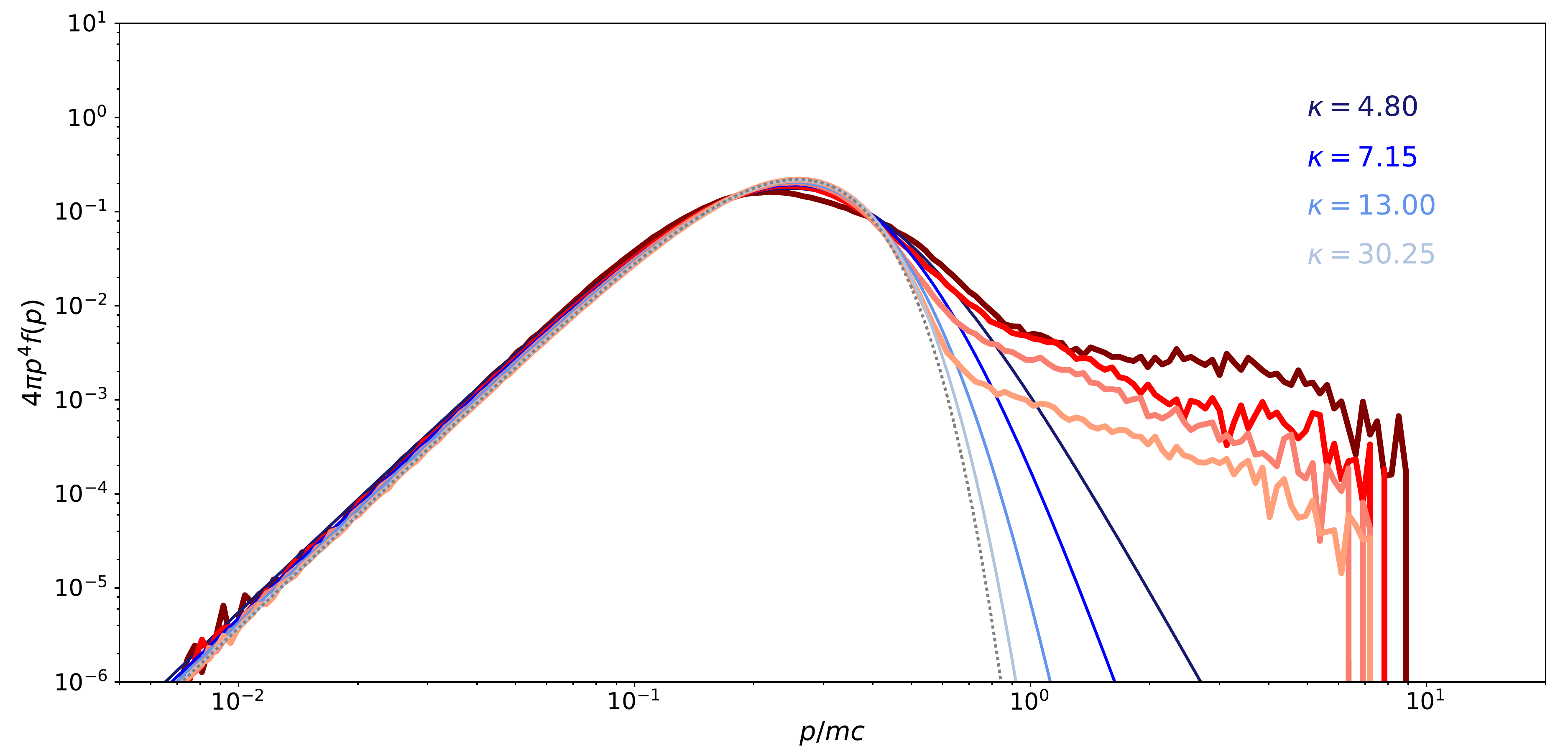}
\caption{{The downstream ion spectra from the regions distributed at different distances behind the shock, which are marked in Fig. 1 by the same color coding as applied in this plot (red lines).} The data is plotted in red: the darker lines correspond to the regions closer, and the lighter lines to the regions farther from the shock. The $\kappa$-distribution fits are represented by blue lines, each fit-line corresponding to the data-line with the same intensity level. The same color coding holds for the $\kappa$ parameters given for each fit (the upper right corner). The Maxwellian fit of the spectrum corresponding to the farthermost downstream region, is plotted by the dotted grey line.}
\end{figure}

In Fig. 3, we show the ion spectra captured at different regions in the downstream. As moving farther from the shock, the $\kappa$-index in the particle distribution increases. This means that the plasma farther from the shock, appears to be closer to its equilibrium state. In the limiting case, an infinite value of $\kappa$ would correspond to the equilibrium case with the Maxwellian distribution.

By tracing the time evolution of the ion spectrum right behind the shock, we find that the value of $\kappa$-index varies in the range $\sim 4-7$. This however holds over the simulation time period, but may be subject to changes over the longer periods, or given the different shock parameters and inclinations relative to the magnetic field.

\section{Conclusions}

The conclusion of this work can be summarized as follows.
\begin{itemize}

\item We showed that the sum of thermal and suprathermal components in the downstream ion spectrum can {theoretically and empirically} be represented by {a} single  $\kappa$ momentum distribution, which is used to describe non-equilibrium plasmas. {We used an approach of non-extensive statistical mechanics to show how the Maxwellian and the suprathermal distribution, that can be related to the particles with constant escape probability in the minimal model of Caprioli et al. (2015), can emerge directly from the generalized entropy.}

\item The spectra closer to the shock imply the non-equilibrium plasma states. Farther from the shock, the plasma settles down, the $\kappa$-index increases, and the $\kappa$-distribution takes the shape of a thermal Maxwellian distribution. The far downstream spectrum is thus composed only of a Maxwellian and a {power law (as in the model of Blasi et al. 2005, or Arbutina \& Zekovi\'c 2021).}

\item The $\kappa$ + modified power-law best fitting procedure {gives} $v_{inj}\sim 1.5-2~v_{sh}$, which implies that ions enter the acceleration process right after the first reflection. The probability of particle being trapped by the reforming barrier, will determine which particles stay only energized and populate the downstream $\kappa$-distribution, and which among them will continue to accelerate through the DSA mechanism. The obtained injection momentum remains nearly the same throughout the downstream.

\end{itemize}

{\bf Acknowledgements.} During the work on this paper the authors were financially supported by the Ministry of Education, Science and Technological Development of the Republic of Serbia through the contract No. {451-03-9/2021-14/200104.} The PIC simulations were run on the cluster JASON of Automated Reasoning Group (ARGO) at the Department of Computer Science, and on a new cluster SUPERAST at the Department of Astronomy, Faculty of Mathematics, University of Belgrade.

\end{document}